\documentclass[10pt,journal,compsoc]{IEEEtran}

\usepackage{xcolor}
\usepackage{color}
\usepackage{amsfonts}       
\usepackage{amsmath}
\usepackage{graphicx}
\usepackage{multirow}
\usepackage{booktabs}
\usepackage{bbding}
\usepackage{ragged2e}
\ifCLASSOPTIONcompsoc
  \usepackage[nocompress]{cite}
\else
  \usepackage{cite}
\fi
\ifCLASSINFOpdf
\else
\fi
\newcommand\MYhyperrefoptions{bookmarks=true,bookmarksnumbered=true,
pdfpagemode={UseOutlines},plainpages=false,pdfpagelabels=true,
colorlinks=true,linkcolor={blue},citecolor={blue},urlcolor={blue},
pdftitle={Bare Demo of IEEEtran.cls for Computer Society Journals},
pdfsubject={Typesetting},
pdfauthor={Michael D. Shell},
pdfkeywords={Computer Society, IEEEtran, journal, LaTeX, paper,
             template}}
\ifCLASSINFOpdf
\usepackage[\MYhyperrefoptions,pdftex]{hyperref}
\else
\usepackage[\MYhyperrefoptions,breaklinks=true,dvips]{hyperref}
\usepackage{breakurl}
\fi
\begin{document}
%
\title{An Audio-Visual Speech Separation Model Inspired by Cortico-Thalamo-Cortical Circuits}
%
%
%
%

\author{Kai Li,~\IEEEmembership{Student Member,~IEEE,}
        Fenghua Xie,
        Hang Chen,
        Kexin Yuan,
        and~Xiaolin Hu,~\IEEEmembership{Senior Member,~IEEE}
\IEEEcompsocitemizethanks{\IEEEcompsocthanksitem K. Li, H. Chen, X. Hu are with the Department of Computer Science and Technology, Institute for Artiﬁcial Intelligence, State Key Laboratory of Intelligent Technology and Systems, BNRist, IDG/McGovern Institute for Brain Research, Tsinghua Laboratory of Brain and Intelligence (THBI), Tsinghua University, Beijing, China. X. Hu is also with the Chinese Institute for Brain Research (CIBR), Beijing, China.
\IEEEcompsocthanksitem F. Xie and K. Yuan are with the Department of Biomedical Engineering, School of Medicine, IDG/McGovern Institute for Brain Research, Tsinghua Laboratory of Brain and Intelligence (THBI), Tsinghua University, Beijing, China. \protect\\
(corresponding authors: Kexin Yuan and Xiaolin Hu.)


}

}

%
%

\markboth{Journal of \LaTeX\ Class Files,~Vol.~14, No.~8, August~2015}%
{Shell \MakeLowercase{\textit{et al.}}: Bare Advanced Demo of IEEEtran.cls for IEEE Computer Society Journals}
%



\IEEEtitleabstractindextext{%
\begin{abstract}
\justifying
Audio-visual approaches involving visual inputs have laid the foundation for recent progress in speech separation. However, the optimization of the concurrent usage of auditory and visual inputs is still an active research area. Inspired by the cortico-thalamo-cortical circuit, in which the sensory processing mechanisms of different modalities modulate one another via the non-lemniscal sensory thalamus, we propose a novel cortico-thalamo-cortical neural network (CTCNet) for audio-visual speech separation (AVSS). First, the CTCNet learns hierarchical auditory and visual representations in a bottom-up manner in separate auditory and visual subnetworks, mimicking the functions of the auditory and visual cortical areas. Then, inspired by the large number of connections between cortical regions and the thalamus, the model fuses the auditory and visual information in a thalamic subnetwork through top-down connections. Finally, the model transmits this fused information back to the auditory and visual subnetworks, and the above process is repeated several times. The results of experiments on three speech separation benchmark datasets show that CTCNet remarkably outperforms existing AVSS methods with considerably fewer parameters. These results suggest that mimicking the anatomical connectome of the mammalian brain has great potential for advancing the development of deep neural networks.
\end{abstract}

}

\maketitle

\IEEEdisplaynontitleabstractindextext

%
\IEEEpeerreviewmaketitle

\ifCLASSOPTIONcompsoc
\IEEEraisesectionheading{\section{Introduction}\label{sec:introduction}}
\else
\section{Introduction}
\label{sec:introduction}
\fi

\IEEEPARstart{H}{uman} have the innate ability to separate various audio signals, such as distinguishing different speakers’ voices or differentiating voices from background noise. This innate ability is known as the ``cocktail party effect" \cite{RN1, RN2}.
Auditory systems analyze the statistical structures of patterns in sound streams (e.g., spectrum or envelope) and can easily identify sounds of interest in sound mixtures \cite{RN78}.
Designing speech separation systems that are as robust as humans has long been a goal in the field of artificial intelligence (AI).

 Researchers have proposed various audio-only speech separation (AOSS) methods \cite{RN18, RN16, RN17} for modeling the cocktail party effect. Although these methods have achieved success in certain idealized settings (e.g., without reverberations or noise), they face significant challenges when addressing more than two speakers (the commonly used PIT training method \cite{RN21} becomes inefficient, if not impossible to use, due to the label permutation problem \cite{RN19}) or when the number of speakers is unknown \cite{RN20, RN22}. 
 
 Previous neuroscience studies have suggested that the human brain typically uses visual information to assist the auditory system in addressing the cocktail party problem \cite{RN72, RN82}. Inspired by this finding, visual information has been incorporated to improve the speech separation performance \cite{RN9, RN10, RN26, RN53, RN118, RN7}. The resulting methods are known as audio-visual speech separation (AVSS) approaches. When visual cues are incorporated, regardless of the number of speakers or whether this number is known a priori, speech separation becomes substantially easier because the system knows who is speaking at any time. In addition, if lip motion can be captured by the system, this additional clue assists in speech processing because it complements information loss in the speech signal in noisy environments \cite{RN72, RN80}. Nevertheless, the abilities of existing AVSS methods are far inferior to the capability of the human brain.

\begin{figure*}[!t]
\centering
\includegraphics[width=0.8\textwidth]{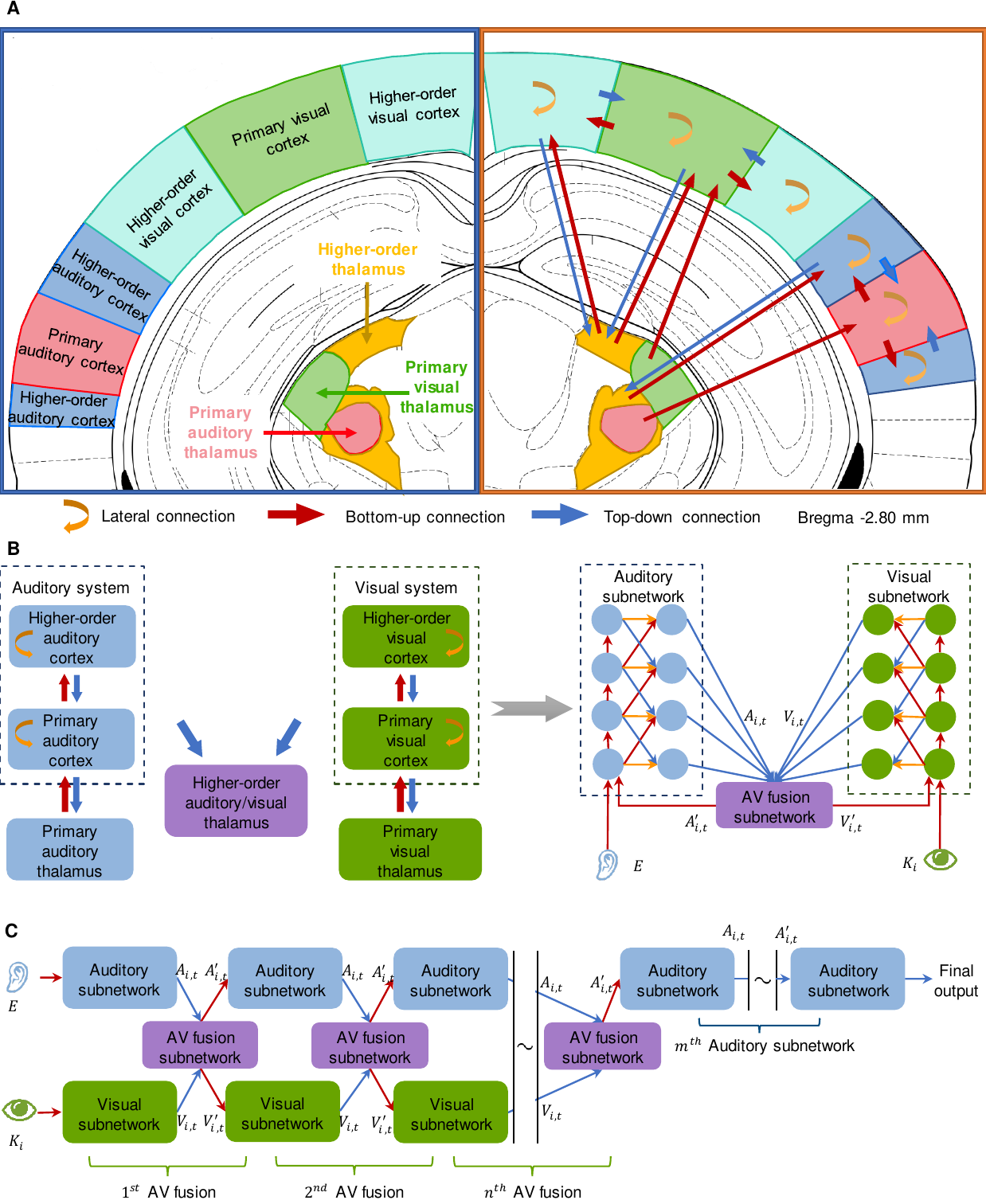}
\caption{Overview of multimodal information flow in the CTCNet model. (A) The multimodal information processing between the auditory cortex and thalamus of the rodent brain is illustrated on the coronal section. The auditory and visual thalamus and cortices and the multimodal thalamus are labeled using different colors. The transmission and integration pathways for different sensory information are delineated on the right side of the figure. The red, blue, and yellow arrows indicate bottom-up, top-down, and lateral connections, respectively. The higher-order sensory thalamic nuclei are involved in intercortical information transmission and integrate auditory and visual inputs. (B) The multimodal information process of the cortico-thalamo-cortical transthalamic connectivity patterns (left) and CTCNet structure (right). In this panel, $\mathbf{E}$ and $\mathbf{K}_i$ denote the audio mixture embedding and the visual features from the auditory and visual modules, respectively; $\mathbf{A}_{i,t}$ and $\mathbf{A'}_{i,t}$ denote auditory features before and after the multi-modal fusion process, respectively; $\mathbf{V}_{i,t}$ and $\mathbf{V'}_{i,t}$ denote visual features before and after the multi-modal fusion process, respectively. (C) The recurrent process of AV fusion in the CTCNet model over time. The red, blue, and yellow arrows indicate bottom-up, top-down, and lateral connections, respectively. In the figure, $n$ represents the number of cycles in the auditory, visual and thalamic subnetworks, and $m$ represents the number of extra cycles in the auditory subnetwork after AV fusion.}
\label{fig1}
\end{figure*}

From a neuroscience perspective, the above AVSS methods simulate supracortical pathways from lower functional areas (e.g., the V1 and A1 areas) to higher functional areas (e.g., the V2 and A2 areas) in primates. 
The structures of most AVSS methods  \cite{RN10, RN26, RN53} are heavily influenced by convolutional neural networks \cite{lecun1989backpropagation}, which were partially inspired by Hubel and Wiesel's findings on cat visual cortex \cite{RN101}.
However, existing AVSS methods overlook the significance of subcortical structures in audio-visual information integration. 

Previous neuroscience studies have revealed that the thalamus, a subcortical brain region, is critical for multimodal integration \cite{RN29, RN40, RN73, RN76, shepherd2021untangling, rouiller2000comparative, keifer2015comparative}. Fig.~\ref{fig1}A illustrates the interaction between thalamic nuclei and different cortical areas in rodent brains. This is a typical diagram in mammals. The hierarchical and complex pathways between thalamic nuclei and different cortices transmit and integrate sensory information, including unimodal thalamocortical pathways (from the primary auditory or visual thalamus to the corresponding primary cortices) and higher-order cortical thalamocortical pathways with multimodal sensory thalamic nuclei. The non-lemniscal or higher-order regions of the auditory thalamus receive anatomical inputs directly from brain regions for visual processing, including the superior colliculus and secondary visual cortex \cite{RN76}, and exhibit excitatory or inhibitory responses to both visual information and auditory inputs \cite{RN41, RN36, RN32}. The auditory cortex receives visual input from the visual thalamus \cite{RN45} and the cortex and integrates auditory and visual information \cite{RN137, RN134}. The complex neural circuits among the sensory thalamus and the cortex \cite{shepherd2021untangling}, especially the transthalamic pathways via the higher-order thalamus, could involve brain processing mechanisms that address the cocktail party effect and thus could motivate novel AVSS methods.

Here, we propose a cortico-thalamo-cortical neural network (CTCNet, Fig.~\ref{fig1}B) for addressing the AVSS task. Our model was inspired by the connections from thalamic to cortical regions involving different sensory modalities, as well as the functions of higher-order sensory thalamic regions that integrate audio-visual information. The CTCNet uses the same hierarchical structure for both the auditory and visual subnetworks. Due to the enormous number of recurrent connections in our model, the sensory information in the auditory and visual subnetworks (corresponding to cortical regions) cycles from lower layers to higher layers while different layers have distinct temporal resolutions, and the auditory and visual information are fused multiple times through a thalamic subnetwork (corresponding to the higher-order thalamus). We conducted extensive experiments and found that the CTCNet outperformed existing methods by a large margin.

\section{Related work}
\label{sec:rw}

\textbf{Audio-only speech separation (AOSS):} Speech separation \cite{RN17, RN16, RN21, RN19} is a widely studied acoustic task that aims to extract different speaker speech signals from the mixed speech signals of multiple speakers. Recently, speech separation performance has been substantially improved due to the development of deep learning techniques. Since it is challenging to estimate the phase in the time-frequency domain, researchers have migrated the speech separation task to the time-domain to avoid this problem and obtained state-of-the-art performance compared to the time-frequency domain, such as DualPathRNN \cite{RN17}, Sepformer \cite{RN18}, etc. However, these methods do not consider the fusion of features at different temporal resolutions. Recently, multi-scale AOSS methods such as AFRCNN\cite{RN46} have achieved a balance between efficiency and separation performance. However, their separation performance in noisy scenarios is still not as good as AVSS methods that leverage visual cues.

\textbf{Audio-visual speech separation (AVSS):} The integration of visual and auditory information for speech separation has sparked considerable interest among researchers \cite{RN9, RN10, RN26, RN53, RN118, RN7, RN54}, principally due to the consonance of such multi-modal neural network integration with the brain's processing modality. In the speech separation tasks, utilizing human facial characteristics to guide the speech separation of the target speaker has been shown to yield impressive separation performance \cite{RN9, RN10, RN26, RN53, RN118, RN7, RN54, martel2023audio}. In more complex acoustic settings, AVSS methods have demonstrated superiority over AOSS methods. However, the efficient fusion of visual and auditory features remains an area of ongoing exploration. Lee et al. \cite{RN9} applied regularization techniques to align visual and auditory information, thereby enhancing the robustness of their models. Wu et al. \cite{RN10} used the time-domain speech separation method Conv-TasNet as a backbone network to design a time-domain multi-modal separation model, and the separation performance was substantially improved compared to the frequency-domain model. Afouras et al. \cite{RN63} employed attention mechanisms to construct correlations between visual and auditory information, effectively improving separation performance. Gao et al. \cite{RN26} applied a multi-task learning framework to jointly learn audio-visual speech separation and cross-modal face-voice embeddings, generalizing well to challenging real-world videos of diverse scenarios. Martel et al. \cite{martel2023audio} achieved better separation performance with less model computation by iteratively processing visual and auditory features. It is a contemporaneous work of this work. Nonetheless, we find that these methods have some common problems. Regarding multi-modal feature fusion, these methods mainly rely on a single temporal scale (e.g., at the smallest temporal scale). However, such processing is insufficient for comprehensive multi-modal fusion. 
In this paper, we present CTCNet, a novel AVSS model that integrates audio and video information at multiple temporal scales. This has led to excellent performance.

\section{The CTCNet-based AVSS model}

\begin{table*}[ht]
\centering
\caption{Some important symbols used in the paper}
\begin{tabular}{c|l}
\toprule
Symbol & Description \\
\midrule
\multicolumn{2}{l}{Overall pipeline} \\
\midrule
$\mathbf{x}$      &  The mixed speech signals with multiple speakers \\
$\mathbf{s}_i$    &  The clean audio of the $i$-th speaker   \\
$\mathbf{\sigma}$ &  The background noise (including environmental noise as well as non-speech meaningful sounds) \\
$\mathbf{P}_i$    &  The visual cues of the $i$-th target speaker \\
$\mathbf{E}$      &  The audio mixture embedding \\
$\mathbf{K}_i$    &  The visual embedding of the $i$-th target speaker \\
$\mathbf{G}_i$    &  The separation network output \\
$\mathbf{M}_{i}$  &  The estimated mask of the $i$-th target speaker \\
$\mathbf{Q}_{i}$  &  The estimated audio features of the $i$-th target speaker \\
\midrule
\multicolumn{2}{l}{CTCNet architecture} \\
\midrule
$\mathbf{A}_{i,t,d}$ & The auditory features of the $i$-th target speaker at the $d$-th downsampling of the $t$-th cycle \\
$\mathbf{V}_{i,t,d}$ & The visual features of the $i$-th target speaker at the $d$-th downsampling of the $t$-th cycle \\
$\mathcal{Q},\mathcal{G}$        & The $1\times1$ convolutional layer \\
$\mathcal{D}$        & The down-sampling layer by using the $\text{stride}=2$ convolutional layer \\
$\mathcal{U}$        & The upsampling layer using the interpolation method \\
$\mathcal{L}_1,\mathcal{L}_2$    & The fully connected layer \\
$\mathcal{F}$ & The fusion strategy by using concatenation or summation \\

\bottomrule
\end{tabular}
\label{tab:symbol}
\end{table*}

\begin{figure*}[!t]
\centering
\includegraphics[width=0.8\textwidth]{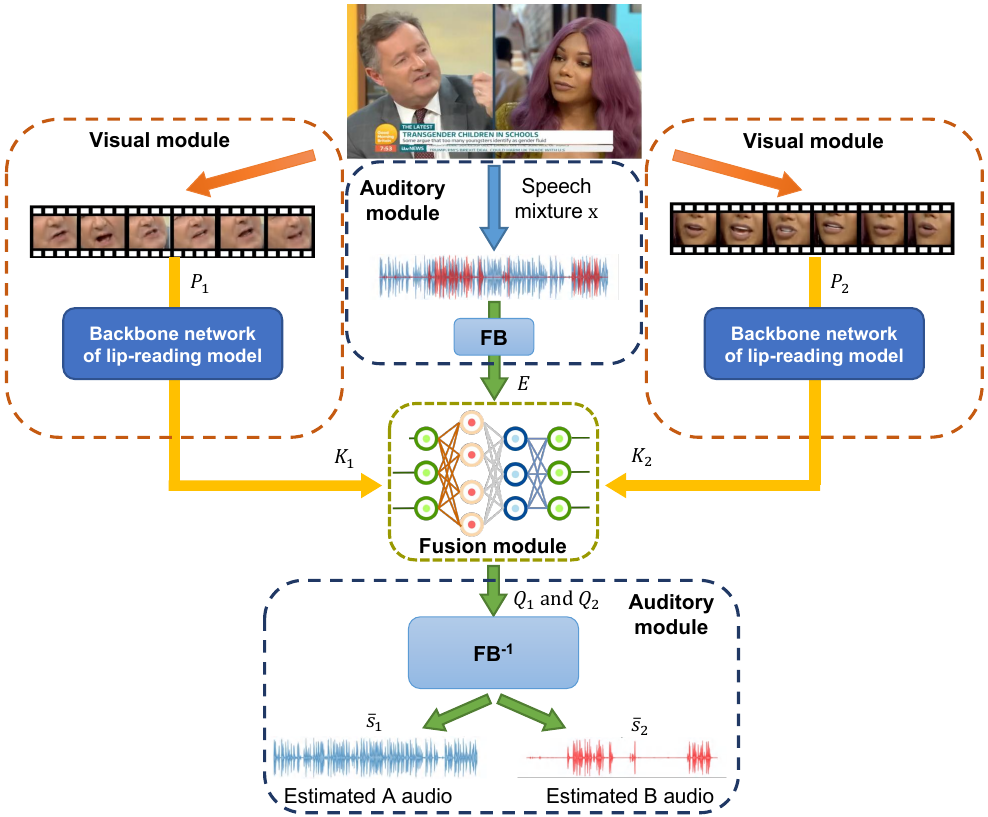}
\caption{The pipeline of our AVSS network. Our network includes an auditory module, a visual module and an AV fusion module. It takes a video with a speech mixture as input and outputs the separated speech of different speakers.}
\label{fig2}
\end{figure*}

\subsection{Overall pipeline} 

Our proposed model includes an \textit{auditory module}, a \textit{visual module} and an \textit{audio-visual (AV) fusion module} (Fig.~\ref{fig2}). The model takes a video containing several people speaking simultaneously as input. The sound is fed into the auditory module to extract auditory features, while the image frames are fed into the visual module to extract visual features. 
%
The auditory module takes auditory features processed by a filter bank (FB) and outputs the results to the AV fusion module.
Then, the fused features from the AV fusion module are converted back to the target speaker's voice by an inverse filter bank (FB$^{-1}$).
The visual module is pre-trained on the lip-reading task. This module extracts visual features and transmits the processed features to the AV fusion module. The AV fusion module was inspired by cortico-thalamo-cortical circuits in the brain, which is described in Section 2.2. The symbols used in Section 3 are summarized in Table~\ref{tab:symbol}.

Specifically, for a given video containing $C$ speakers, we assume that the speech mixture $\mathbf{x}\in R^{1\times T_a}$ in the video consists of linearly superimposed voices $\mathbf{s}_i\in R^{1\times T_a}$ of different speakers,
\begin{equation}
    \mathbf{x}=\sum_{i=1}^C \mathbf{s}_i + \mathbf{\sigma},
\end{equation}
where $\mathbf{\sigma}\in R^{1\times T_a}$ denotes background noise (including environmental noise as well as non-speech meaningful sounds) and $T_a$ denotes the audio length. The aim is to estimate $\mathbf{s}_i$ according to $\mathbf{x}$ by using the visual cues $\mathbf{P}_i\in R^{H\times W\times T_v}$ of the target speaker $i$, where $H$, $W$, and $T_v$ denote the height, width, and number of the image frames, respectively.


The auditory module first uses $N$ one-dimensional kernels with length $L$ and stride $\left \lfloor L/2 \right \rfloor $ to perform one-dimensional same-mode convolution\footnote{The 1D convolution and transposed convolution are both implemented in Pytorch with nn.Conv1d and nn.ConvTranspose1d.} with  $\mathbf{x}$, resulting in an audio mixture embedding $\mathbf{E}\in R^{N\times T_e}$, where $T_e=\frac{T_a-L}{\left \lfloor L/2 \right \rfloor} + 1$. The $N$ kernels are called FB \cite{RN16, RN17}. The visual module accepts the image frames $\mathbf{P}_i\in R^{H\times W\times T_v}$ and extracts the features $\mathbf{K}_i\in R^{C_v\times T_v}$ using the backbone network in the lip-reading model (see Section 2.4), where $C_v$ denotes the number of output channels of the backbone network. The AV fusion module receives $\mathbf{K}_i$ and $\mathbf{E}$ as inputs. After processing, the final output $\mathbf{G}_i\in R^{N\times T_e}$ of the AV fusion module passes through a fully connected layer followed by a nonlinear activation (ReLU) function to estimate the target speaker’s mask $\mathbf{M}_{i}\in R^{N\times T_e}$.

Then, the auditory module reconstructs the target speaker’s voice $\mathbf{\bar{s}}_i\in R^{1\times T_a}$ according to the mask $\mathbf{M}_{i}$. First, the target speaker’s auditory features $\mathbf{Q}_i = (\mathbf{E}\odot \mathbf{M}_{i})\in R^{N\times T_e}$ are extracted using element-wise multiplication $\odot$ between $\mathbf{M}_{i}$ and $\mathbf{E}$. Next, the estimated signal $\mathbf{\bar{s}}_i\in R^{1\times T_a}$ is reconstructed using $\mathbf{Q}_i$ through a FB$^{-1}$ \cite{RN16, RN17}, which is implemented by  one-dimensional same-mode transposed convolution with $N$ input channels, kernel size $L$ and stride $\left \lfloor L/2 \right \rfloor$. 

Unlike the AOSS methods, most existing AVSS methods do not need to know the number of speakers beforehand. This is attributed to the additional visual inputs in the model, allowing the model to extract relevant audio from the speakers identified visually. During the training and inference processes, the CTCNet, like other AVSS methods \cite{RN10, RN53, RN54}, essentially predicts one speaker’s voice at a time (although in implementation, we can predict all speaker’s voices in parallel). 

\subsection{The CTCNet}
The fusion module, CTCNet, consists of an \textit{auditory subnetwork}, a \textit{visual subnetwork} and a \textit{thalamic subnetwork} (Fig.~\ref{fig1}B). The auditory and visual subnetworks have hierarchical structures that correspond to auditory and visual pathways in the cortex, respectively. The lower layers have higher temporal resolutions, while the higher layers have lower temporal resolutions. The thalamic subnetwork has a single-layer structure corresponding to the higher-order AV thalamus. This subnetwork consolidates the auditory and visual information with the highest temporal resolution.

The CTCNet works as follows (Fig.~\ref{fig1}B, right). First, the auditory and visual subnetworks take the auditory features $\mathbf{E}\in R^{N\times T_e}$ and visual features $\mathbf{K}_i\in R^{C_v\times T_v}$ extracted by the auditory and visual modules (Fig.~\ref{fig2}) as inputs, respectively. These subnetworks process this information in a bottom-up manner until the information arrives at the top layers of the subnetworks using down-sampling scale $2^{(d-1)}$, obtaining multi-scale visual and auditory features ($\{\mathbf{A}_{i,t,d}\in R^{N\times \frac{T_e}{2^{(d-1)}}}| d=1,...,S\}$ and $\{\mathbf{V}_{i,t,d}\in R^{C_v\times \frac{T_v}{2^{(d-1)}}}| d=1,...,S\}$), where $2^{(d-1)}$ denotes the scale at which visual and auditory features are down-sampled; $d$ denotes the number of down-sampling; $S$ is the number of layers in the auditory and visual subnetworks. Second, the unimodal features from adjacent layers are resized to the same size by down-sampling (bottom-up), up-sampling (top-down) or copying (lateral) operations, followed by a fusion operation ($1\times 1$ convolution) at all temporal resolutions:
\begin{equation}
\begin{aligned}
    \mathbf{A}_{i,t,d}=\mathcal{Q} (Concat(\mathcal{D}(\mathbf{A}_{i,t,(d-1)}), \mathbf{A}_{i,t,d}, \mathcal{U}(\mathbf{A}_{i,t,(d+1)}))),\\ 
    \mathbf{V}_{i,t,d}=\mathcal{Q} (Concat(\mathcal{D}(\mathbf{V}_{i,t,(d-1)}), \mathbf{V}_{i,t,d}, \mathcal{U}(\mathbf{V}_{i,t,(d+1)}))).
\end{aligned}
\end{equation}
Where $\mathcal{Q}$ denotes the $1\times1$ convolutional layer to compress channels of features concatenated, $\mathcal{D}$ stands for the down-sampling layer by using the $\text{stride}=2$ convolutional layer and $\mathcal{U}$ represents the upsampling layer using the interpolation method. Note that each layer receives information from the lower, higher and same-order layers, resembling sensory integration in different cortical areas, which is governed by bottom-up, top-down and lateral synapses in unimodal sensory pathways \cite{RN112, RN113}. 
Third, the fused features at all temporal resolutions in the auditory and visual subnetworks, denoted by $\mathbf{A}_{i,t}\in R^{N\times T_e}$ and $\mathbf{V}_{i,t}\in R^{C_v\times T_v}$ where the subscript $t$ denotes the $t$-th cycle, are projected to the thalamic subnetwork, mimicking the large number of neural projections from different cortical areas to the higher-order thalamus \cite{RN115, RN92}:
\begin{equation}
\begin{aligned}
    \mathbf{A}_{i,t}=\mathcal{G}(Concat(\mathbf{A}_{i,t,1}, \mathcal{U}(\{\mathbf{A}_{i,t,d}| d=2,...,S\}))),\\ 
    \mathbf{V}_{i,t}=\mathcal{G}(Concat(\mathbf{V}_{i,t,1}, \mathcal{U}(\{\mathbf{V}_{i,t,d}| d=2,...,S\}))),
\end{aligned}
\end{equation}
where $\mathcal{G}$ denotes the 1x1 convolutional layer to compress channels of features concatenated.

The thalamic subnetwork works as follows: $\mathbf{V}_{i,t}$ and $\mathbf{A}_{i,t}$ are firstly resized to the same sizes as $\mathbf{A}_{i,t}$ and $\mathbf{V}_{i,t}$, respectively; then, the features are concatenated or summed in the feature dimension; finally, the fully connected layers $\mathcal{L}_1(\cdot)$ and $\mathcal{L}_2(\cdot)$ are used to obtain the features $\mathbf{V'}_{i,t}\in R^{C_v\times T_v}$ and $\mathbf{A'}_{i,t}\in R^{N\times T_e}$ respectively. These steps can be formalized as
\begin{equation}
    \mathbf{V'}_{i,t}=\mathcal{L}_1(\mathcal{F}(\mathbf{V}_{i,t}, \mathcal{U}(\mathbf{A}_{i,t}))), \mathbf{A'}_{i,t}=\mathcal{L}_2(\mathcal{F}(\mathbf{A}_{i,t}, \mathcal{U}(\mathbf{V}_{i,t}))),
\end{equation}
where $\mathcal{F}(\cdot)$ represents either concatenation or summation. After the multimodal fusion process, the new fused features ($\mathbf{A'}_{i,t}$ and $\mathbf{V'}_{i,t}$) are transmitted back to the auditory and visual subnetworks, and the above process is repeated. 

The entire process resembles the dynamic behavior of biological systems with feedback connections. Fig.~\ref{fig1}C illustrates this process by demonstrating the working process over time. To reduce computational costs, after several fusion cycles, we remove the visual subnetwork and cycle information through only the auditory subnetwork to obtain the final output $\mathbf{G}_i$.

\begin{figure*}[!t]
\centering
\includegraphics[width=0.8\textwidth]{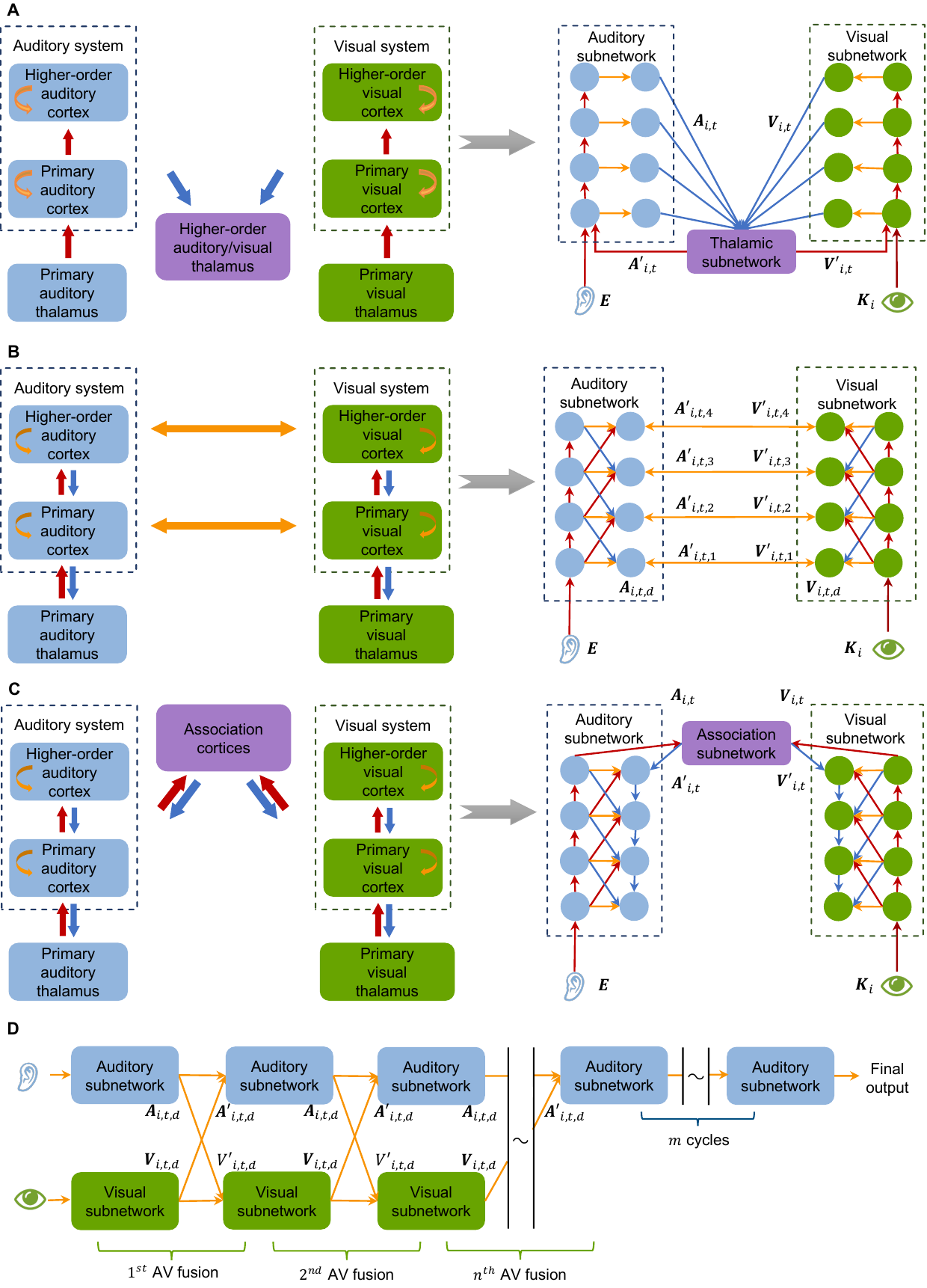}
\caption{Alternative models for AV fusion. (A) The DFTNet model was obtained by removing cross-connections between adjacent layers in the CTCNet model, which corresponds to a CTC diagram in the brain without top-down connections along unimodal cortical pathways. (B) The CCNet model was obtained by directly connecting the auditory and visual subnetworks in CTCNet, which was inspired by the cross-connections between the auditory and visual pathways in the brain, where $\mathbf{A}_{i,t,d}$ and $\mathbf{A'}_{i,t,d}$ denote auditory features from different temporal resolutions before and after the multi-modal fusion process, respectively; $\mathbf{V}_{i,t,d}$ and $\mathbf{V'}_{i,t,d}$ denote visual features from different temporal resolutions before and after the multi-modal fusion process, respectively. (C) The CACNet model was obtained by moving the AV fusion subnetwork to the top layers of the auditory and visual subnetworks in CTCNet. (D) The recurrent process of AV fusion in CCNet over time, where the structure of the auditory subnetwork with m cycles is the same as that of the auditory subnetwork of CTCNet, with inputs and outputs $A_{i,t}$ and $A^{'}_{i,t}$, respectively. The notations are the same as those in Fig.~\ref{fig1}. The yellow arrows in D indicate that only visual and auditory features in the same levels in the two subnetworks are fused.}
\label{fig3}
\end{figure*}

The auditory and visual subnetworks share a similar structure to our previously proposed AOSS model A-FRCNN \cite{RN46}, which is obtained by exploring the order of information flow in a fully recurrent convolutional neural network (FRCNN) \cite{liao2016bridging} through extensive experiments. In the context of AV fusion, the projection target of the top-down connections from all layers of the subnetworks is interpreted as the multimodal thalamus. Note that the FRCNN could have many variants and different variants may fit different modalities. However, to make the CTCNet simple enough, we use the same structure (but with different parameters) for the two subnetworks without an attempt to explore more appropriate variants for the visual subnetwork.

\subsection{Control models}
\textbf{DFTNet}: In addition to projections from all layers of the auditory and visual subnetworks to the thalamic subnetwork, another property of the CTCNet is local fusion between layers in the unimodal sensory subnetworks. To investigate the importance of this property in addressing our task, a control model was constructed by removing the bottom-up and top-down connections between adjacent layers (Fig.~\ref{fig3}A). This model is known as the direct feedback transcranial network (DFTNet). This network has the same workflow as the CTCNet except that the second step is absent.

\textbf{CCNet}: Previous neuroscience studies have suggested the existence of interconnections between the visual and auditory pathways \cite{RN92, RN93} (Fig.~\ref{fig3}B). For example, using neuroanatomical tracers, researchers observed connections between the visual and auditory cortex \cite{RN137}. We were interested in whether similar interconnections between the auditory and visual subnetworks in our model could help in addressing the speech separation task. Thus, we designed an alternative AV fusion module by connecting the auditory and visual subnetworks directly and removing the thalamic subnetwork. The resulting model is known as the cortico-cortical neural network (CCNet, Fig.~\ref{fig3}B right). The CCNet fuses auditory and visual features ($\mathbf{A}_{i,t,d}\in R^{N\times \frac{T_e}{2^{(d-1)}}}$ and $\mathbf{V}_{i,t,d}\in R^{C_v\times \frac{T_v}{2^{(d-1)}}}$) in the same layer $d=[1,..,S]$, through lateral connections to obtain the new fused features ($\mathbf{A'}_{i,t,d}\in R^{N\times \frac{T_e}{2^{(d-1)}}}$ and $\mathbf{V'}_{i,t,d}\in R^{C_v\times \frac{T_v}{2^{(d-1)}}}$), which are then sent back to the auditory and visual subnetworks, respectively. The above process is formulated by
\begin{equation}
\begin{aligned}
    \mathbf{V'}_{i,t,d}=\mathcal{L}_{1,d}(F(\mathbf{V}_{i,t,d}, R(\mathbf{A}_{i,t,d}))),\\ \mathbf{A'}_{i,t,d}=\mathcal{L}_{2,d}(F(\mathbf{A}_{i,t,d}, R(\mathbf{V}_{i,t,d}))).
\end{aligned}
\end{equation}
Then this process is removed after several fusion cycles, and the information only cycles through the auditory subnetwork (Fig.~\ref{fig3}D). See \textbf{Appendix A} for details.

\textbf{CACNet}: Previous neuroscience studies have also revealed that multimodal fusion occurs in certain higher functional areas, including the frontal cortex, occipital cortex and temporal association cortex \cite{RN73, RN87, RN88} (Fig.~\ref{fig3}C). For example, in humans, the posterior superior temporal sulcus (STS) is involved in AV integration \cite{RN133}. We were interested in whether this high-level integration of sensory information between our designed auditory and visual subnetworks was necessary for speech separation. Thus, we designed another AV fusion module called the cortico-association-cortical network (CACNet). This model differs from the CTCNet mainly in the position of the AV fusion subnetwork, located in the top layers of the unimodal subnetworks. To ensure that the fused features ($\mathbf{A'}_{i,t}$ and $\mathbf{V'}_{i,t}$) propagated downward in the unimodal subnetworks, we introduced vertical top-down connections in these subnetworks (Fig.~\ref{fig3}C, right). The CACNet workflow is described in \textbf{Appendix A}. Similar models to CACNet have been proposed for other AV tasks \cite{RN89, RN90, RN91, dai2022binaural, yu2023brain}; however, existing models are usually feedforward models without recurrent connections.

\subsection{Lip-reading pre-training}
The CTCNet and control models use the pre-training lip-reading model CTCNet-Lip to extract visual features. 
Following the existing lip-reading method MS-TCN \cite{RN55}, the CTCNet-Lip model consists of a backbone network for extracting features from the image frames and a classification subnetwork for word prediction. The backbone network includes a 3D convolutional layer and a standard ResNet-18 network. Given multiple grayscale images of the target speaker's lip region in the input video $\mathbf{P}_i\in R^{H\times W\times T_v}$, we first convolve the frames with $J$ 3D kernels with size $5\times 5\times 1$ and spatial stride size $2\times 2\times 1$ to obtain the features $\mathbf{F}_v\in R^{J\times H'\times W'\times T_v}$, where $H'$ and $W'$ denote the spatial dimensions. This step performs feature extraction and compression of the input multiple grayscale images through a 3D convolutional operation. Then, each frame in $\mathbf{F}_v$  passes through the ResNet-18 network, resulting in a $C_v$-dimensional embedding. Finally, the $T_v$ frames compose an embedding matrix $\mathbf{K}_i\in R^{C_v\times T_v}$.

The classification subnetwork takes $\mathbf{K}_i$ as input for word prediction. In contrast to MS-TCN, which uses a temporal convolutional network (TCN) for word prediction, we use a model with the same structure as CTCNet's visual subnetwork, whose output has the same dimensions as $\mathbf{K}_i$. The output is averaged in the time dimension, resulting in a $C_v$-dimensional vector. A fully connected layer and a softmax layer are used for word prediction.

After lip-reading pre-training, the backbone network of the CTCNet-Lip is fixed for extracting visual features for CTCNet, and the classification subnetwork is discarded.

\subsection{Loss function for training}
The scale-invariant source-to-noise ratio (SI-SNR) \cite{RN58} between the estimated and original signals is used as the loss function for CTCNet and control models. Each speaker's SI-SNR is defined as
\begin{equation}
    \text{SI-SNR}=20\times \log_{10}(\frac{||\mathbf{\alpha}\cdot \mathbf{s}||}{||\mathbf{\bar{s}}-\mathbf{\alpha}\cdot \mathbf{s}||}),
\end{equation}
where
\begin{equation}
    \mathbf{\alpha}=\frac{\bar{\mathbf{s}}^T\mathbf{s}}{\mathbf{s}^T\mathbf{s}},
\end{equation}
$\mathbf{s}\in R^{1\times T_a}$and $\bar{\mathbf{s}}\in R^{1\times T_a}$ denote the original and estimated audio, respectively, which are normalized to zero mean.

\section{Experiments}
\subsection{Datasets}
Following previous studies, we constructed speech separation datasets by mixing single speech from three commonly used AV datasets, LRS2 \cite{RN47}, LRS3 \cite{RN48}, and VoxCeleb2 \cite{RN49}. The speakers in the test set of these datasets did not overlap with those in the training and validation sets. Speeches and image frames were obtained from the videos using the FFmpeg\footnote{https://ffmpeg.org} tool. The length of each speech was 2 seconds, and the sampling rate was 16 kHz. The videos were synchronized with speech and sampled at 25 FPS. To be easier for training and testing, we used a common construction of datasets in the speech separation domain, setting the number of two speakers. 

The LRS2 dataset contains thousands of BBC video clips. It has three folders: \texttt{Train}, \texttt{Validation} and \texttt{Test}. We constructed the LRS2-2Mix dataset from it, which contains two-speaker speech mixtures.  The training and validation sets were constructed by randomly selecting different speakers from the \texttt{Train} and \texttt{Validation} folders, followed by mixing speeches with signal-to-noise ratios between -5 dB and 5 dB. This resulted in an 11-hour training set and a 3-hour validation set. We used the same test set\footnote{https://github.com/afourast/avobjects} as in previous works \cite{RN54,RN26,RN9,RN63}.

The LRS3 dataset includes thousands of spoken sentences in TED and TEDx videos. It has two folders: \texttt{Trainval} and \texttt{Test}. We constructed the LRS3-2Mix dataset from the LRS3 dataset. The training and validation sets were constructed by randomly mixing different speakers' audios from the \texttt{Trainval} folder, subject to the constraint that the speakers in the training set and validation set did not overlap. The mixing settings were the same as LRS2-2Mix.  This resulted in a 28-hour training set and a 3-hour validation set. We used the same test set\footnote{https://github.com/caffnet} as in previous works \cite{RN9,RN26,RN54}.

The VoxCeleb2 dataset contains over 1 million quotes from 6,112 celebrities extracted from YouTube videos. It has two folders: \texttt{Dev} and \texttt{Test}. Due to computational resource constraints, we only selected 5\% of the training and validation data in the \texttt{Dev} folder of VoxCeleb2 to construct an AV dataset named VoxCeleb2-2Mix. The training and validation sets contain 56-hour and 3-hour mixture speeches, respectively. The mixing settings were same as LRS2-2Mix. We used the same test set\footnote{http://dl.fbaipublicfiles.com/VisualVoice} as in previous works \cite{RN26,RN9}.

In the created three datasets: LRS2-2Mix, LRS3-2Mix and VoxCeleb2-2Mix, the first and third contain noisy and multilingual speeches, while the second contains clean and monolingual (English) speeches.
Following the MIT license, our scripts for generating the training and validation datasets were provided through an open-source repository\footnote{https://github.com/JusperLee/LRS3-For-Speech-Separation}.

\subsection{Implementation details}
In the auditory module, we set $L=21$ and $N=512$ in the FB and FB$^{-1}$. In the visual module, we pre-trained the lip-reading model CTCNet-Lip on the LRW \cite{RN56} dataset with the same settings as MS-TCN \cite{RN55}. 
In the AV fusion module, the auditory and visual subnetworks of CTCNet have a similar structure, with the following differences. In the auditory subnetwork, we used standard 1-D convolutional layers with kernel size 5 and 512 channels, followed by global layer normalization (GLN) \cite{RN16}. In the visual subnetwork, we found that standard 1-D convolutional layers with smaller kernel sizes and fewer channels (3 and 64) worked better. In addition, we used batch normalization \cite{RN138} instead of GLN. The number of layers $S$ in the auditory and visual subnetworks was set to 5. In the unimodal fusion of the auditory and visual subnetworks, the bottom-up connections were realized by the standard 1-D convolutional layer with kernel size 5 and stride size 2; the top-down connections were realized by nearest neighbors interpolation; the lateral connections were realized by copying the feature maps. In the thalamic subnetwork, we set the number of channels in the $1\times 1$ convolutional layer to 576.

The control models, DFTNet, CCNet and CACNet, used the same settings as described above. DFTNet (Large) increased the number of channels in each 1-D convolutional layer of the auditory subnetwork from 512 to 768.

When training the CTCNet-Lip model, we used the same experimental setup as for MS-TCN \cite{RN55}. Specifically, CTCNet-Lip was trained using the LRW1000 dataset. During training, we applied a cosine annealing scheduler for 100 epochs and used the Adam optimizer for model optimization. The initial learning rate of the model was set to $3\times 10^{-4}$, while a weight decay of $1\times 10^{-4}$ and a batch size of 32 were applied. As for the loss function, we chose the standard cross-entropy loss function.

When training the CTCNet and control models, we used the AdamW \cite{RN139} optimizer with a weight decay of $1\times 10^{-1}$. We used $1\times 10^{-3}$ as the initial learning rate and halved the learning rate when the validation set loss did not decrease in 5 successive epochs. Gradient clipping was used to limit the maximum $l_2$-norm value of the gradient to 5. The training lasted a maximum of 200 epochs, with early stopping applied. All models were trained on 8$\times$ NVIDIA Tesla V100 32G GPUs. 

The proposed method was implemented in Python and PyTorch. All code and scripts for reproducing the results will be released after the acceptance of the paper.

\subsection{Evaluation metrics}
We used the scale-invariant signal-to-noise ratio improvement (SI-SNRi) and signal-to-noise ratio improvement (SDRi) to evaluate the quality of the separated speeches. These metrics were calculated based on the scale-invariant signal-to-noise ratio \cite{RN58} and signal-to-noise ratio \cite{RN59}:
\begin{equation}
    \text{SI-SNRi}=\text{SI-SNR}(\bar{\mathbf{s}}, \mathbf{s})-\text{SI-SNR}(\mathbf{s}, \mathbf{x}),
\end{equation}
\begin{equation}
    \text{SDRi}=\text{SDR}(\bar{\mathbf{s}}, \mathbf{s})-\text{SDR}(\mathbf{s}, \mathbf{x}),
\end{equation}
where $\mathbf{x}\in R^{1\times T_a}$, $\mathbf{s}\in R^{1\times T_a}$and $\mathbf{\bar{s}}\in R^{1\times T_a}$ denote the audio mixture, original audio and estimated audio, respectively.

\section{Results}

\subsection{Hyperparameter setting}

\begin{table}[!t]
\centering
\caption{Comparison of different AV fusion strategies. $\uparrow$ indicates that higher values are better.}
\label{tab1}
\begin{tabular}{cc|cc}
\toprule
$n$                  & Strategy & SDRi (dB) $\uparrow$ & SI-SNRi (dB) $\uparrow$ \\
\midrule
\multirow{2}{*}{1} & Concat   & 13.2                 & 12.9                        \\
                   & Sum      & 13.4                 & 13.2                \\
\midrule
\multirow{2}{*}{2} & Concat   & 13.3                 & 13.0                \\
                   & Sum      & 13.4                 & 13.1                \\
\midrule
\multirow{2}{*}{3} & Concat   & 13.6                 & 13.3                \\
                   & Sum      & \textbf{13.8}                 & \textbf{13.6}                     \\
\midrule
\multirow{2}{*}{4} & Concat   & 13.5                 & 13.2                \\
                   & Sum      & 13.6                 & 13.3                \\
\bottomrule
\end{tabular}
\end{table}

\textbf{AV fusion strategies}: We used the LRS2-2Mix dataset to examine the influence of two important parameters on the CTCNet: the number of AV fusion cycles ($n$ in Fig.~\ref{fig1}C) and the multimodal feature fusion strategy (summation or concatenation). The number of cycles in the auditory subnetwork after AV fusion ($m$ in Fig.~\ref{fig1}C) was set to 5. In general, we observed that the model was not sensitive to either $n$ or the fusion strategy (Table~\ref{tab1}). With both fusion strategies, we noted a slightly increasing trend following a decreasing trend with increasing $n$. 
By considering the tradeoff between the results and the computational cost, we chose $n=3$ and the summation fusion strategy in our other experiments. The summation strategy is biologically plausible considering that neurons in the brain can actually integrate inputs from different modalities in an additive manner, which has been confirmed in many brain areas \cite{komura2005auditory, kobayasi2013audiovisual, knopfel2019audio}.

\begin{table}[!t]
\centering
\caption{Performance of CTCNet with different values of $m$ in Fig.~\ref{fig1}C, with $n$ fixed at 3.}
\label{tab2}
\begin{tabular}{c|cccc}
\toprule
$m$  & SDRi (dB) $\uparrow$ & SI-SNRi (dB) $\uparrow$ \\
\midrule
1  & 11.8      & 11.4        \\
5  & 13.5      & 13.2          \\
13 & 14.6      & 14.3   \\
17 & 14.8      & 14.5\\
21 & \textbf{14.9}      & \textbf{14.6}\\
\bottomrule
\end{tabular}
\end{table}

\textbf{Number of auditory cycles}: Then, we examined the effect of the number of cycles in the auditory subnetwork after AV fusion ($m$ in Fig.~\ref{fig1}C) on the separation performance. The performance improved as $m$ increased (Table~\ref{tab2}). However, this improvement is somewhat limiting. More importantly, as the number of cycles $m$ increases, the training and inference time of the model will rise significantly, and accordingly, the complexity of the model will also increase. Our model design balances pursuing higher performance and maintaining model complexity and operational efficiency. Therefore, $m$ was set to 13 only in the experiment comparing the CTCNet with existing methods. In all other experiments, $m$ was set to 5.

\begin{table}[h]
\centering
\caption{Comparison of different weight sharing strategies. ``\XSolid" indicates that the subnetwork shares weights across time steps. ``\Checkmark" indicates that the visual subnetwork shares weights across time steps.}
\label{tab3}
\begin{tabular}{cc|cccc}
\toprule
Auditory                  & Visual                    & SDRi (dB) $\uparrow$ & SI-SNRi (dB) $\uparrow$ \\
\midrule
\XSolid    & \XSolid    & 13.3      & 13.1    \\
\Checkmark & \XSolid    & \textbf{13.6}      & \textbf{13.3}       \\
\Checkmark & \Checkmark & 13.5      & 13.2         \\
\bottomrule
\end{tabular}
\end{table}

\textbf{Weight sharing}: CTCNet is a recurrent neural network. This model needs to be converted to a feedforward network by unfolding the network through time for training and testing (Fig.~\ref{fig1}C). The resulting feedforward network uses the same set of weights in different cycles that correspond to the same structure at various time steps. However, if different sets of weights are used, the model performance may be greatly enhanced since considerably more parameters are introduced. We compared the model results with and without weight sharing and observed that the difference was negligible (Table~\ref{tab3}). We, therefore, applied weight sharing in all other experiments.

\begin{table}[h]
\centering
\caption{Comparison of CTCNet performance on the LRS2-2Mix dataset with /without pre-training the lip-reading model.}
\label{tab4}
\begin{tabular}{c|cccc}
\toprule
Module Type & SDRi (dB) $\uparrow$ & SI-SNRi (dB) \\
\midrule
CTCNet (without pre-training) & 4.5           & -2.8                \\
CTCNet (with pre-training)    & \textbf{13.5} & \textbf{13.2} \\
\bottomrule
\end{tabular}
\end{table}

\textbf{Lip-reading pre-training}: A lip-reading model, CTCNet-Lip, was used to preprocess the image frames to obtain the visual features. Our experimental results indicated that this model needed to be pre-trained on a lip-reading dataset (Table~\ref{tab4}). When the CTCNet-Lip model was trained from scratch, CTCNet overfitted the training set and obtained poor results on the test set. pre-training is a standard technique in deep learning when the training set is not large and diverse enough.

\subsection{CTCNet could separate speeches from mixtures}

\begin{figure*}[!t]
\centering
\includegraphics[width=0.8\textwidth]{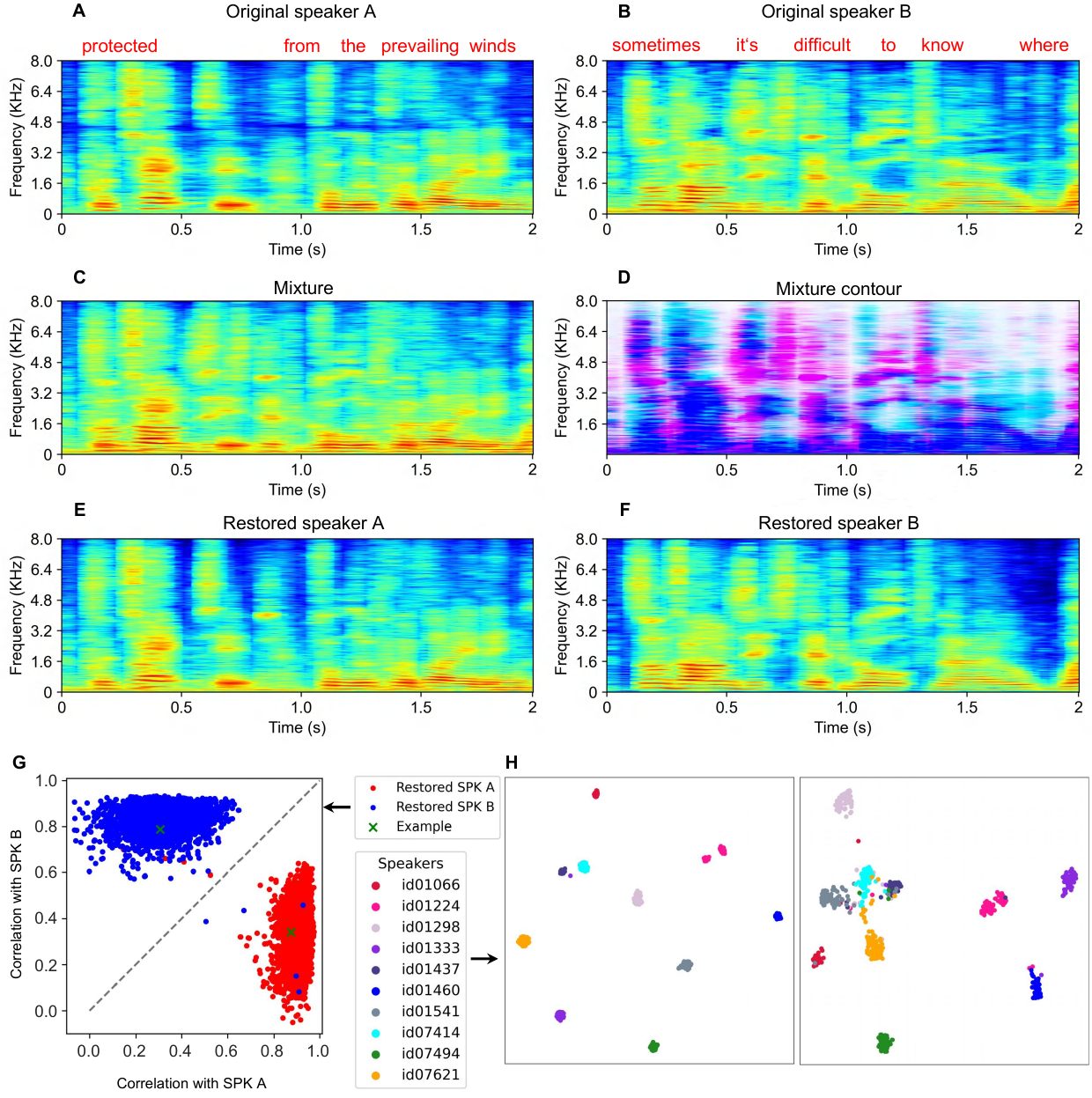}
\caption{Visualization of speech separation results obtained by the CTCNet. (A) and (B) Two example target speech spectrograms. The corresponding textual content was labeled in the corresponding positions above the spectrograms. (C) The mixture speech spectrogram. (D) The mixture contour. (E) and (F) The two restored speech spectrograms. (G) Scatter plot of correlation for all test examples (N=3,000). Each point indicates the correlation between the spectrograms of the separated speeches and the target speeches. The point ``$\times$" indicates the correlation for the example of the displayed spectrograms. (H) Visualization of the speaker identity of the target speeches (left) and separated speeches (right) by the t-SNE method. Each point corresponds to the speaker's embedding from the x-vector.}
\label{fig4}
\end{figure*}

To visualize the results of the CTCNet, we randomly selected one speech mixture in the LRS2-2Mix test set. We used it as input to the CTCNet to estimate two speakers’ speeches. The spectrograms of the original speeches, mixture, and restored speeches are shown in Fig.~\ref{fig4}. The mixture contour (Fig.~\ref{fig4}D) indicated that the spectral features of the two sources were highly overlapping, demonstrating the difficulty of restoring the sources. However, the CTCNet obtained excellent results on this example (Fig.~\ref{fig4}E and \ref{fig4}F). To quantify the quality of results, we calculated the correlation between the two separated spectrograms and target spectrograms for all test examples. The correlations were scattered above or below the diagonal (Fig.~\ref{fig4}G), indicating that the separated spectrograms had higher correlations with the target spectrograms and lower correlations with the irrelevant spectrograms.

In addition, to examine the speaker identity of the separated speeches, we input speech mixtures from the VoxCeleb-2Mix test set into CTCNet and fed the outputs into the speaker recognition model x-vector \cite{RN110} to extract speaker features. The x-vector model was trained by SpeechBrain \cite{RN103} on the VoxCeleb2 \cite{RN49} dataset. We were interested in whether the auditory features output from the CTCNet retained speaker identity. One-hour mixtures from 10 speakers were randomly selected. We used the x-vector sixth layer features to obtain the speaker features and applied t-distributed stochastic neighbor embedding (t-SNE) \cite{RN96} to visualize the distribution of the speaker features (Fig.~\ref{fig4}H, right). The features of different speakers tend to form distinct clusters, similar to the distribution of the target speech (Fig.~\ref{fig4}H, left). This demonstrated that CTCNet can preserve speaker identity when separating speech.

\subsection{Visual information helps speech separation}

\begin{table*}[!t]
\centering
\caption{Comparison between CTCNet and the control models on the LRS2-2Mix (test set). CTCNet (Audio-only) is a CTCNet containing only the auditory subnetwork. DFTNet (Large) is a larger model than DFTNet. $\downarrow$ indicates that lower values are better; $\uparrow$ indicates that higher values are better.}
\label{tab5}
\begin{tabular}{c|ccccc}
\toprule
Module Type         & SDRi (dB) $\uparrow$ & SI-SNRi (dB) $\uparrow$ & Params (M) $\downarrow$   \\
\midrule
CTCNet (Audio-only) & 10.1          & 9.4                     & 6.3          \\
DFTNet              & 13.1          & 12.8                    & \textbf{3.6} \\
DFTNet (Large)      & 13.1          & 12.9                    & 7.6          \\
CCNet               & 10.9          & 10.4                  & 18.6         \\
CACNet              & 10.2          & 9.9                  & 7.1          \\
\midrule
CTCNet              & \textbf{13.5} & \textbf{13.2}  & 7.0          \\
\bottomrule
\end{tabular}
\end{table*}

We investigated the effect of the visual information on the separation performance with the LRS2-2Mix dataset. The CTCNet used the default settings. For comparison, we tested a model with only the auditory subnetwork, and this subnetwork was cycled $n+m=8$ times. The AVSS model outperformed the AOSS model by a large margin (Table~\ref{tab5}). The AVSS model obtained relative improvements of 33.7\%, 40.4\% over the AOSS model in terms of the SDRi and SI-SNRi metrics.

These results suggest that visual information significantly improves the separation performance, which is consistent with previous neuroscience \cite{RN72, RN73} and AVSS studies \cite{RN9, RN10, RN26, RN53, RN47}.

\subsection{CTCNet outperformed the control models}
We compared the results obtained by CTCNet and three control models, namely, DFTNet, CCNet, and CACNet, on the LRS2-2Mix dataset (Table~\ref{tab5}). DFTNet was designed to investigate the usefulness of unimodal fusion in the unimodal subnetworks of CTCNet, while CCNet and CACNet were designed to compare different AV fusion methods. The default settings were used for CTCNet. Similar settings were used for the three control models.

We made the following observations. First, DFTNet obtained worse results than CTCNet. Since DFTNet has half the number of parameters of CTCNet, to ensure a fair comparison, we designed a larger version of DFTNet, denoted by DFTNet (Large) in Table~\ref{tab5}, with a comparable number of parameters to CTCNet by increasing the number of convolutional channels. However, increasing the number of parameters led to a negligible performance improvement. This result suggests the importance of fusing features in different layers of the unimodal subnetworks.

Second, CTCNet outperformed CCNet by a large margin, indicating that direct fusion between the auditory and visual subnetworks is not as effective as fusion with a thalamic subnetwork. In CCNet, each layer in the auditory and visual subnetworks does not receive as much multiresolution information during each AV fusion cycle as in CTCNet.

Third, CTCNet outperformed CACNet by a large margin, indicating that AV fusion in the bottom layers of the auditory and visual subnetworks is more effective than AV fusion in the top layers. The auditory and visual features in the top layer may be too coarse to provide useful information for guiding speech separation.

\begin{figure*}[!t]
\centering
\includegraphics[width=0.7\textwidth]{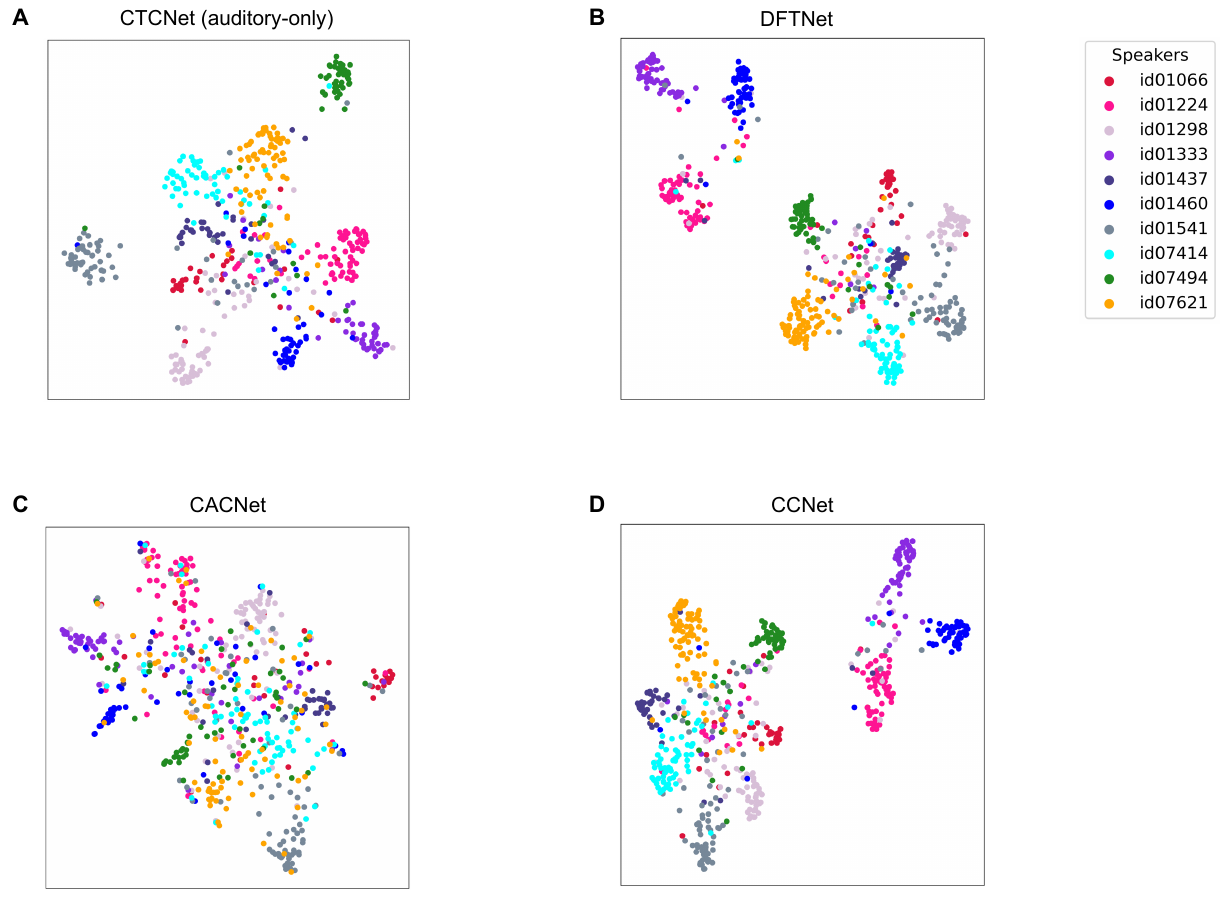}
\caption{Visualization of the speaker identity of the separated speech output by different control models using the t-SNE method. The same convention is used as in Fig.~\ref{fig4}H.}
\label{fig5}
\end{figure*}

Finally, according to the results of the x-vector model, CTCNet showed a better clustering effect for the speaker features than the control models (compare Fig.~\ref{fig4}H and Fig.~\ref{fig5}), indicating better performance of CTCNet for keeping the speaker identity.

\subsection{CTCNet outperformed existing methods}

\begin{table*}[!t]
\centering
\caption{Comparison between CTCNet and existing methods on the LRS2-2Mix, LRS3-2Mix and VoxCeleb2-2Mix datasets. ``*" indicates that the method uses ground-truth phases for speech signal reconstruction. ``-" indicates that the results were not reported in the original papers. ``$\dagger$" indicates results were obtained by us with the aid of the Asteroid toolkit \cite{RN64} under the MIT license. ``$\flat$" indicates results were obtained by using the codes provided by the original authors. ``Train/Val" denotes the data scales of training and validation sets. For these methods, the real-time factor (RTF) indicates the estimated time consumption per second on the GPU (NVIDIA 1080Ti) and CPU (Intel(R) Xeon(R) Silver 4210 CPU @ 2.20 GHz).}
\label{tab6}
\begin{tabular}{c|cc|cc|cc|ccc}
\toprule
\multirow{2}{*}{Model} & \multicolumn{2}{c}{LRS2-2Mix} & \multicolumn{2}{c}{LRS3-2Mix} & \multicolumn{2}{c}{VoxCeleb2-2Mix} & \multirow{1}{*}{Params} & \multirow{1}{*}{RTF} & \multirow{1}{*}{RTF } \\
                       & SI-SNRi      & SDRi    & SI-SNRi    & SDRi   & SI-SNRi    & SDRi    &    (M)                        &  (GPU, ms)  &        (CPU, s)                     \\
\midrule
\multicolumn{10}{c}{Audio-only speech separation (AOSS) methods}                                                                                                                                                       \\
\midrule
DPCL++$^\dagger$\cite{RN19}                & 3.3           & 4.3         & 5.8           & 6.2     & 2.1              & 2.5      & 13.6  &  233.85                   & 0.16                       \\
uPIT$^\dagger$\cite{RN21}                 & 3.6           & 4.8   & 6.1           & 6.4    & 2.4              & 2.8    & 92.7   & 108.24                    & 0.29                       \\
Conv-TasNet$^\dagger$ \cite{RN16}          & \textbf{10.3} & \textbf{10.7} & 11.1          & 11.4      & 6.9              & 7.5   & 5.6         &   \textbf{15.28}             & 0.19                       \\
SuDORM-RF$^\dagger$  \cite{RN62}           & 9.1           & 9.5     & 12.1          & 12.3    & 6.5              & 6.9    &  \textbf{4.6} &     27.86         & \textbf{0.14}              \\
A-FRCNN$^\dagger$    \cite{RN46}           & 9.4           & 10.1  & \textbf{12.5} & \textbf{12.8}  & \textbf{7.8}     & \textbf{8.2}   & 6.3      &           61.16        & 0.58                       \\
\midrule
\multicolumn{10}{c}{Audio-visual speech separation (AVSS) methods}                                                                                                                                             \\
\midrule
The Conversation \cite{RN53}      & -             & -         & -             & -        & -             & 8.9   &  -  & -                        & -                      \\
LWTNet  \cite{RN54}               & -             & 10.8    & -             & 4.8     & -                & -    &  -  & -                       & -                       \\
AV-ConvTasNet \cite{RN10} & 12.5$^\flat$             & 12.8$^\flat$    & 11.2$^\flat$             & 11.7$^\flat$        & 9.2$^\flat$                & 9.8$^\flat$       & 16.5          &      60.3        & 1.46                       \\
VisualVoice \cite{RN26}           & 11.5$^\flat$          & 11.8$^\flat$   & 9.9$^\flat$           & 10.3$^\flat$   & 9.3              & 10.2     & 77.8           &      130.2       & 3.60                       \\
CaffNet-C  \cite{RN9}            & -             & 10.0    & -             & 9.8      & -              & 7.6         & -          & -                 &  -                          \\
CaffNet-C*   \cite{RN9}           & -             & 12.5    & -        & 12.3     & -            & -         & -             & -                &  -                      \\
Thanh-Dat   \cite{RN63}           & -             & 11.6  & -             & -    & -                & -        & -           & -                  &  -                          \\
AVLIT-8 \cite{martel2023audio}& 12.8$^\flat$ & 13.1$^\flat$ & 13.5$^\flat$ & 13.6$^\flat$ & 9.4$^\flat$    & 9.9$^\flat$  & \textbf{5.6} &   \textbf{53.4}            & \textbf{1.38}       \\
CTCNet                 & \textbf{14.3} & \textbf{14.6} & \textbf{17.4} & \textbf{17.5} & \textbf{11.9}    & \textbf{13.1} & 7.0  &   109.9            & 1.60       \\
\bottomrule
\end{tabular}
\end{table*}

\begin{table*}[!t]
\centering
\caption{Comparison of the influence of different models with various pre-trained models on the separation performance. ``Pre-trained" denotes the pre-trained lip-reading model. ``Acc" denotes the recognition accuracy on the lip-reading dataset LRW. The fourth and fifth columns show the separation results.}
\label{tab7}
\begin{tabular}{cc|ccc}
\toprule
Model       & Pre-trained & Acc (\%) $\uparrow$      & SDRi (dB) $\uparrow$     & SI-SNRi (dB) $\uparrow$    \\
\midrule
VisualVoice & MS-TCN     & \textbf{85.3} & 11.5          & 11.8        \\
CTCNet      & MS-TCN     & \textbf{85.3} & 13.9          & 14.2       \\
VisualVoice & CTCNet-Lip & 84.1          & 11.6          & 11.9         \\
CTCNet      & CTCNet-Lip & 84.1          & \textbf{14.3} & \textbf{14.6}  \\
\bottomrule
\end{tabular}
\end{table*}

We compared CTCNet with existing AOSS and AVSS methods on three datasets: LRS2-2Mix, LRS3-2Mix and VoxCeleb2-2Mix (Table~\ref{tab6}). In this experiment, CTCNet used the default settings, with the exception of $m=13$. 

Existing audio-visual speech separation (AVSS) methods used different training and validation sets. For instance, on LRS2-2Mix, the method in \cite{RN63} used 29 hours data for training and 1 hour data for validation, while the method in \cite{RN26} used 11 hours data for training and 3 hours data for validation. In addition, many methods did not report the size of the training and validation sets in the original papers (e.g., \cite{RN9}). It is therefore hard to compare these methods in a strictly fair setting. Fortunately, all these methods used the same test sets on every dataset. Please note that the sizes of the training and validation sets for CTCNet are the same as for the existing AVSS method, as shown in Table~\ref{tab6}. The methods in \cite{RN118} and \cite{RN7} cannot be compared because they use different training and test sets. Please note that only three methods (AVLIT \cite{martel2023audio}, AV-ConvTasNet \cite{RN10} and VisualVoice \cite{RN26}) compared in Table~\ref{tab6} have complete metric values because their source codes are publicly available. The other AVSS methods compared in Table 6 have not been open-sourced.

For the audio-only speech separation (AOSS) methods, we used audio from the training and test sets consistent with CTCNet on the three datasets to obtain the results in Table~\ref{tab6}.

From Table~\ref{tab6}, we made the following observations. First, among the AOSS methods, time-domain methods (Conv-TasNet \cite{RN16}, SudoRM-RF \cite{RN62} and A-FRCNN \cite{RN46} exhibited consistently stronger performance than time-frequency domain methods (DPCL++ \cite{RN19}, uPIT \cite{RN21}. This result explains why we developed CTCNet in the time domain. Second, on the noise-free dataset LRS3-2Mix, existing AVSS methods obtained worse separation results than the AOSS method A-FRCNN \cite{RN46}; however, on the noisy and multilingual datasets VoxCeleb2-2Mix and LRS2-2Mix, some existing AVSS methods (VisualVoice \cite{RN26} and CaffNet-C \cite{RN9}) yielded substantially better results than the AOSS methods. These results indicate that the addition of visual information in existing AVSS methods is helpful in separating degraded and complex speech mixtures. Third, CTCNet achieved considerably better results than existing AVSS methods on three datasets. 

Tabel~\ref{tab6} also lists the amount of trainable parameters and inference time for each method. Because the backbone network in the lip-reading model was fixed after pre-training, its parameters are not counted in the total number of parameters of CTCNet. Without this visual feature extraction network, CTCNet had one order of magnitude fewer trainable parameters than existing AVSS methods. With this visual feature extraction network, CTCNet had 18.2M parameters, which were still three times fewer than VisualVoice

In terms of inference speed, CTCNet was twice as fast as VisualVoice on CPU, and about 16\% faster on GPU. Although AVLIT-8 was more efficient than VisualVoice, its separation quality was low.

These results indicate the superiority of CTCNet on devices with limited computational resources.

\subsection{Effect of changing the lip-reading model}
Since VisualVoice \cite{RN26} used a different lip-reading pre-training model than CTCNet, it was unknown if the enhanced performance of CTCNet was due to the new lip-reading model. Thus, we investigated the effectiveness of different lip-reading pre-training models (MS-TCN \cite{RN55}  and CTCNet-Lip) on the separation performance of VisualVoice and CTCNet on the LRS2-2Mix test set (Table~\ref{tab7}). For VisualVoice the pre-trained lip-reading models were fine-tuned after pre-training, as described in the original paper \cite{RN26}. For CTCNet, the pre-training models were not fine-tuned. Usually, fine-tuning could lead to better results. But to save the training cost we did not do that for CTCNet. 

With both MS-TCN and CTCNet-Lip, CTCNet obtained better separation results than VisualVoice, indicating that the performance gap between CTCNet and VisualVoice was not due to their different pre-trained lip-reading models. Surprisingly, although CTCNet-Lip showed worse performance than MS-TCN on the lip-reading task (84.1\% versus 85.3\% accuracy), it showed better speech separation performance when combined with CTCNet. 
This result might have occurred because CTCNet-Lip can fuse features at different scales and obtain more robust information. However, MS-TCN lacks the ability to fuse multi-scale information. As shown in many previous studies \cite{RN62, RN46, li2022efficient}, multi-scale information fusion is important for speech separation. However, CTCNet-Lip has a U-Net structure, which aims to restore features to the maximum temporal resolution. This structure is usually advantageous for dense prediction tasks such as speech separation \cite{li2022efficient} and image segmentation \cite{ronneberger2015u, zhou2018unet++}, but may not be optimal for coarse prediction tasks such as lip-reading. Analogous to image classification, in lip-reading, every frame is required to output only one label. MS-TCN is more suitable for this task because its feedforward CNN architecture is suitable for acquiring global features, which is advantageous for the lip-reading task.

\subsection{CTCNet outperformed VisualVoice in real-world scenarios}
We were interested in the performance of different AVSS methods in real-world scenarios. We noted that VisualVoice\footnote{https://github.com/facebookresearch/VisualVoice} \cite{RN26} was the open-source method in Table~\ref{tab6}. Thus, we applied CTCNet and VisualVoice to various YouTube videos containing background noise or reverberations, such as debates and online meetings. Because these videos do not have ground-truth separated audio, we could evaluate the results only qualitatively. CTCNet clearly produced higher-quality separated audio than VisualVoice (Supplementary Video 1).

\subsection{Speech separation improves automatic speech recognition}
Speech separation should help downstream tasks such as automatic speech recognition (ASR). We were interested in the potential improvement if CTCNet was applied to speech mixtures before these mixtures were input into ASR systems. In this experiment, the Google Cloud Speech-to-Text API\footnote{https://cloud.google.com/speech-to-text} was used to recognize speech in the LRS2-2Mix test set, as shown in Table~\ref{tab8}. Without speech separation, the word error rate (WER) was 84.91\%. When CTCNet was applied for speech separation, the WER decreased to 24.82\%. This WER is significantly lower than the previously reported WER achieved after applying SOTA methods \cite{RN9, RN26, martel2023audio, RN10} for speech separation. This finding demonstrates the improved performance of CTCNet over existing AVSS methods in downstream applications.

\begin{table}[]
\centering
\caption{The automatic speech recognition results and separation performance on the LRS2 dataset. The ``$*$" indicates that the phase information of the ground truth and the estimated amplitude information were used to recover the audio signal. We utilized the same automatic speech recognition API as CaffNet-C.}
\label{tab8}
\begin{tabular}{lll}
\toprule
Methods       & SI-SNRi (dB) $\uparrow$ & WER (\%) $\downarrow$  \\
\midrule
Mixture       & -       & 84.91     \\
Ground-truth  & -       & 17.74     \\
CaffNet-C \cite{RN9}     & 10.0    & 35.08     \\
CaffNet-C$*$ \cite{RN9}   & 12.5    & 32.96     \\
VisualVoice \cite{RN26}   & 11.8    & 34.45     \\
AV-ConvTasNet \cite{RN10} & 12.5    & 31.43     \\
AVLIT-8 \cite{martel2023audio}       & 12.8    & 31.85     \\
\midrule
CTCNet        & \textbf{14.4}    & \textbf{24.82}     \\
\bottomrule
\end{tabular}
\end{table}

\section{Discussion and Conclusion}
Multimodal processing occurs in various sensory cortical areas, including the frontal cortex, occipital cortex, temporal association cortex and posterior STS \cite{RN73, RN87, RN88}, which perform higher-level functions than unimodal cortical areas. These higher-level multimodal association areas have motivated several AV processing methods \cite{RN9, RN26, RN118, RN7}. These methods usually fuse the outputs of auditory and visual subnetworks in the top layer, such as CACNet in Fig.~\ref{fig3}C. There is also considerable evidence that the auditory and visual cortices have direct neural projections \cite{RN92, RN93}, indicating that AV processing also occurs in unimodal cortical areas. This fact motivated us to construct the CCNet for addressing AVSS tasks in this study (Fig.~\ref{fig3}B).

However, CACNet and CCNet ignore the functions of the thalamus, a subcortical substrate that also performs AV processing. The thalamus in human and other animal brains is an essential hub that integrates the brain’s functional networks. The thalamus is strongly connected to cortical functional networks and interacts homogeneously with multiple sensory networks, thus participating in various cognitive functions \cite{RN29, RN40, RN73, shepherd2021untangling, rouiller2000comparative, keifer2015comparative}. These facts motivated us to propose the CTCNet for addressing AVSS tasks. A key feature of the proposed model is the thalamus-like AV fusion subnetwork, which receives inputs from all layers of the auditory and visual subnetworks, fuses the information, and sends the fused information back to the unimodal subnetworks. Though this subnetwork has a very simple single-layer structure, it has enabled CTCNet to achieve considerably better results than CACNet and CCNet, as well as previous state-of-the-art AVSS methods.


CTCNet, CACNet and CCNet are task-oriented models that are not completely accurate representations of the brain structure. The better performance of CTCNet suggests the importance of a thalamus-like module under the deep learning framework when performing AVSS tasks. However, it does not indicate that higher-level multimodal association areas or interconnections between auditory and visual cortical areas are unimportant for addressing the cocktail party problem. A better AI model for AVSS task could be found by simulating AV fusion functions in all brain regions.

\ifCLASSOPTIONcompsoc
  \section*{Acknowledgments}
\else
  \section*{Acknowledgment}
\fi

This work was supported by the National Key Research and Development Program of China (grant 2021ZD0200301), the National Natural Science Foundation of China (grants U19B2034, 62061136001 and 61836014), and the Tsinghua-Toyota Joint Research Fund.

\ifCLASSOPTIONcaptionsoff
  \newpage
\fi



\bibliographystyle{IEEEtran}
\bibliography{references}
%



%
\begin{IEEEbiography}[{\includegraphics[width=1in,height=1.25in,clip,keepaspectratio]{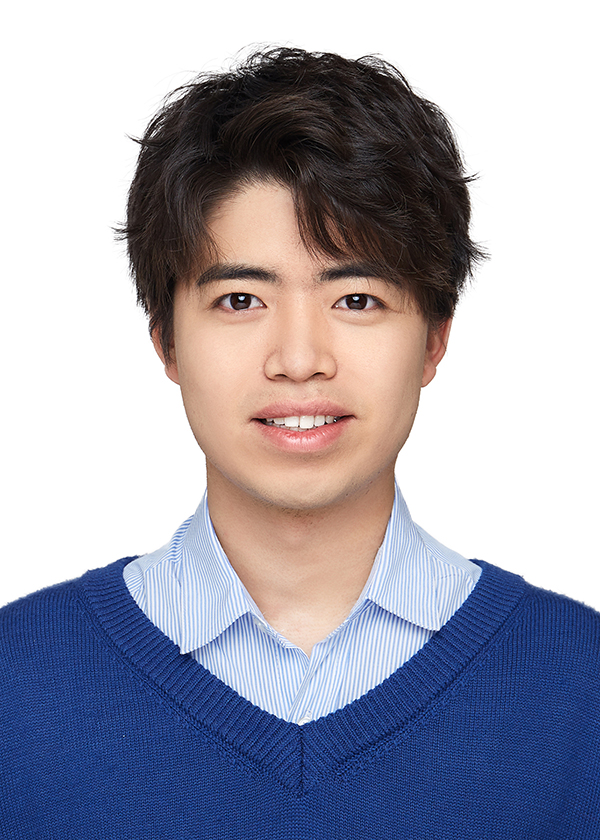}}]{Kai Li} received the B.E. degree from Qinghai University, China, in 2020. He is currently pursuing the M.S. degree with the Department of Computer Science and Technology, Tsinghua University, China. His research interests are primarily on audio processing and audio-visual learning, especially speech separation.
\end{IEEEbiography}

\begin{IEEEbiography}[{\includegraphics[width=1in,height=1.25in,clip,keepaspectratio]{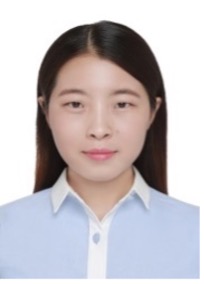}}]{Fenghua Xie} received the B.E. degree in Biomedical Engineering from Beihang University, China, in 2016 and the Ph.D degree in Biomedical Engineering from Tsinghua University, China, in 2022. She is currently a Postdoctor fellow in School of Medicine, Tsinghua University, China. Her research interests include sensory processing, loops between thalamus and cortices, and top-down neural mechanisms.
\end{IEEEbiography}

\begin{IEEEbiography}[{\includegraphics[width=1in,height=1.25in,clip,keepaspectratio]{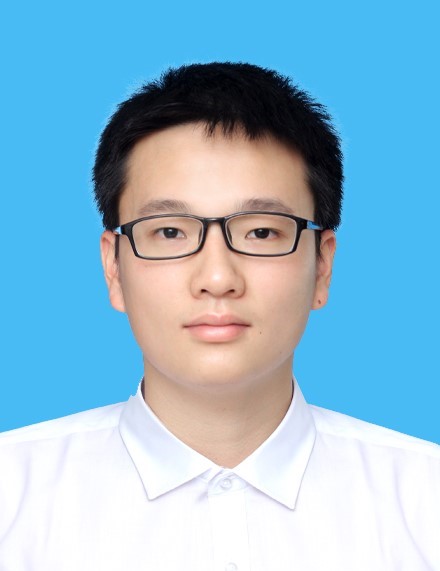}}]{Hang Chen} received the B.S. degree in Physics from Tsinghua University, China, in 2020. He is currently pursuing the Ph.D. degree with the Department of Computer Science and Technology, Tsinghua University, China. His research interests include computer vision and multi-modal learning.
\end{IEEEbiography}

\begin{IEEEbiography}[{\includegraphics[width=1in,height=1.25in,clip,keepaspectratio]{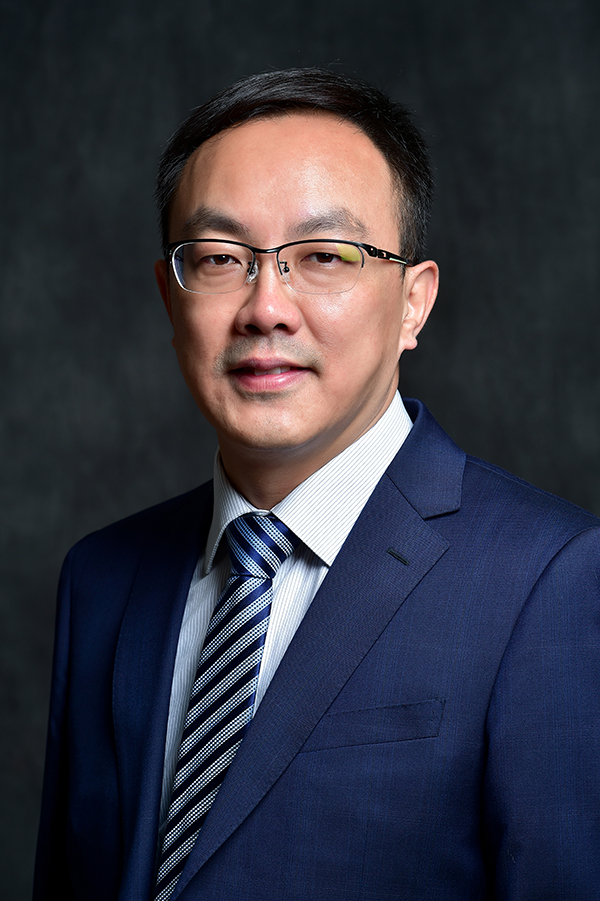}}]{Kexin Yuan} received a B.S. degree in Biochemistry from the Yantai University, Shandong, China, in 2001, and a Ph.D. degree in Neuroscience from the Institute of Biophysics, Chinese Academy of Sciences, Beijing, in 2006. He received his postdoctoral training in Neuroscience from the University of California at San Francisco \& Berkely, USA, from 2006-2012. He is currently an Associate Professor at the Department of Biomedical Engineering, School of Medicine, Tsinghua University, Beijing, China. His research interests mainly focus on the neural mechanisms underlying the translation from sensory inputs to behavioral outputs in animals.
\end{IEEEbiography}

\begin{IEEEbiography}[{\includegraphics[width=1in,height=1.25in,clip,keepaspectratio]{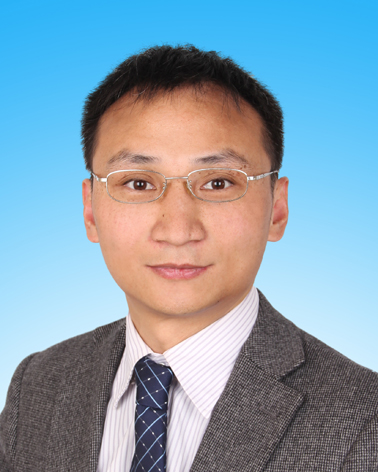}}]{Xiaolin Hu}
(S’01-M’08-SM’13) received B.E. and M.E. degrees in automotive engineering from the Wuhan University of Technology, Wuhan, China, in 2001 and 2004, respectively, and a Ph.D. degree in automation and computer-aided engineering from the Chinese University of Hong Kong, Hong Kong, in 2007. He is currently an Associate Professor at the Department of Computer Science and Technology, Tsinghua University, Beijing, China. His current research interests include deep learning and computational neuroscience. At present, he is an Associate Editor of the IEEE Transactions on Pattern Analysis and Machine Intelligence, IEEE Transactions on Image Processing, and Cognitive Neurodynamics. Previously he was an Associate Editor of the IEEE Transactions on Neural Networks and Learning Systems.
\end{IEEEbiography}







\end{document}